\documentclass[sigplan,screen,sigconf]{acmart}

\usepackage{amsmath}
\usepackage{graphicx}
\usepackage{listings}

\setcopyright{rightsretained}
\acmDOI{}
\acmISBN{}
\acmConference[LATTE '21]{1st Workshop on Languages, Tools, and Techniques for Accelerator Design}{April 15, 2021}{Virtual, Earth}

\author{Nick Brown}
\affiliation{
  \institution{EPCC at the University of Edinburgh}
  \country{UK}
}           
           
\begin{document}

\title{Application specific dataflow machine construction for programming FPGAs via Lucent}

\maketitle

\section{Introduction and background}

Enabling programmers to write code in C or C++, High Level Synthesis (HLS) has been a very significant step forwards in the programmability of Field Programmable Gate Arrays (FPGAs). Synthesising to the Hardware Description Language (HDL) level, the maturity of such tools means that these transformations will likely result in correctly functioning FPGA designs, however algorithms that are still based on Von Neumann (CPU or GPU) approaches are seldom fast. Often there are numerous steps required to tune such algorithms for FPGAs, which can be time consuming and require specialist knowledge, in some cases resulting in thousands of times performance difference between the first FPGA version and the final optimised code but requiring significant programmer effort.

An important consideration is thus how the community can enable programmers to write codes for FPGAs that are \emph{fast by construction}. There have been a number of efforts here, for instance OpenACC for FPGAs \cite{openacc}, but the programmer is still using an imperative languages which is built on Von Neumann foundations. Even DaCe \cite{dace}, which promotes a data centric programming environment via a graph abstraction is based around Python (for instance it still has variables) and aims to target a wide variety of hardware. Whilst there is much to be said for a single framework targeting multiple architectures, due to the fact that FPGAs are based on entirely different hardware foundations, then there is potentially value in exploring approaches which are optimised for that specific hardware. Put simply, it is our hypothesis that an abstraction that is entirely suited towards dataflow and without any Von Neumann baggage, will potentially enable the development of fast by construction codes for FPGAs more effectively.

\section{Application specific dataflow machines}

Dataflow itself has a long history, not least in the 1980s and 1990s with the development of general purpose dataflow machine hardware \cite{manchester}. This was based, in part at-least, on the observation that the dataflow paradigm enables massive amounts of raw concurrency and encourages programmers to focus on the movement of data to keep compute fed. Whist such general purpose dataflow machine hardware was ultimately not successful, it is our view that the fixed general purpose nature of these architectures caused many of the problems. Crucially, the reconfigurable nature of FPGAs offer an opportunity for dataflow machines to be entirely bespoke and tuned to the application in question.

\begin{figure}[h]
\centering
\includegraphics[scale=0.61]{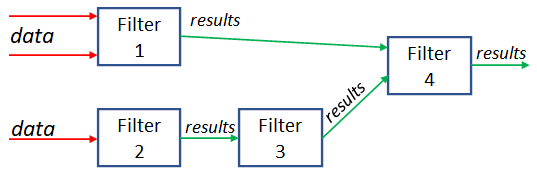}
\caption{Illustration of programmer's dataflow model}
\label{fig:dataflow_architecture}
\end{figure}

It is our view that approaching FPGA programming of constructing \emph{application specific dataflow machines} delivers performance. With the programmer's explicit execution model being that of data streams connecting independent \emph{filters} transforming input data to output data (as illustrated in Figure \ref{fig:dataflow_architecture}), the execution of statements does not follow a strict ordering defined in code but instead is driven by the availability of data with many facets executing concurrently. This enables the programmer to focus on \emph{keeping data continually moving}, and unlike fixed general purpose dataflow machines of history, reconfigurability means that we have control over granularity and filters can be as simple or complex as needed by the application. It is possible in HLS to manually adopt such an execution model using streams and dataflow regions if the programmer is careful, and in \cite{nekbone} it was demonstrated that this approach promotes high performance, with a key question being what support from the programming languages is most helpful?

If one subscribes to the view that developing high performance codes for FPGAs is akin to writing application specific dataflow machines, then effectively the programmer is looking for a language which enables them to describe the construction of such a dataflow machine most effectively. 

\section{Lucent: The child of Lucid}

Arguably one of the more well known dataflow languages targeted at general purpose dataflow machines was Lucid \cite{lucid}. Initially developed in the 1970s, Lucid is a declarative language where the programmer describes the desired results of computations without explicitly saying how this is to be achieved. One can view the original Lucid similarly to a functional language, but one in which the time is a first class concern (these differences are in contrast to other dataflow approaches such as RVC-CAL \cite{cal}). 

It is these two facets, the declarative nature and time as a first class concern, that we believe make Lucid especially useful as a foundation for developing application specific dataflow machines. The declarative nature of the language means that there is a significant amount of high level information which the compiler can use to then infer the most appropriate implementation. Furthermore, making time a first class concern is also crucially important as this forces the programmer to consider the evolution of data from one point in time to the next. Little wonder then that other languages have been based on Lucid, including Lustre \cite{lustre} which was used to program hardware circuits for FPGAs in the early 1990s \cite{lustre-hw} via HDL. Whilst Lustre has enjoyed continued development, the main focus is around developing electronic circuitry for embedded systems, with the language becoming popular for real-time control critical software. This is true of many such approaches, and by contrast our focus is the porting of HPC kernels, enabling the programmer to readily accelerate their scientific algorithms via FPGAs. 



We have developed a programming language called Lucent which is based on the ideas presented in Lucid, but targeted towards HPC programmers who are developing dataflow machines for FPGAs. Designed as a research vehicle to explore these concepts and their applicability to FPGAs, Lucent builds on the foundational concepts of Lucid to present a modern language for 2021 capable of implementing a range of complex algorithms. Our compiler transforms the programmer's Lucent source code to HLS C++ device code (for Xilinx architectures) and generates the required host OpenCL driver code and configuration files for compilation via Xilinx Vitis. 

It is not realistic to describe the entire language in this short position paper, so instead we use examples to highlight some key features and how these encourage the development of fast by construction dataflow codes. Listing \ref{lst:simple_addition} illustrates a very simple addition kernel, summing pairs of numbers. This can be viewed as a single dataflow filter, with two inputs and an output. There is no such thing as a variable in Lucent (as that indicates data at rest, and if at rest then it is not working for us!) Consequently \emph{a} and \emph{b} are both input streams of integers, and \emph{mykernel} is an output stream of type integer. The expression \emph{mykernel = a + b} declares that the value of the output stream at any point in time (what we call a \emph{time quanta} in Lucent) is that of the integer in the \emph{a} stream added to the value in the \emph{b} stream. At each time quanta new data elements are streaming into the kernel via \emph{a} and \emph{b}, added together and the result streamed out of the filter, before moving onto the next time quanta.

\begin{lstlisting}[frame=lines,caption={Simple example of integer stream addition},label={lst:simple_addition}, numbers=left]
external mykernel:int(a:int, b:int) where:
    mykernel=a + b
\end{lstlisting}

\begin{figure}[h]
\centering
\includegraphics[scale=0.65]{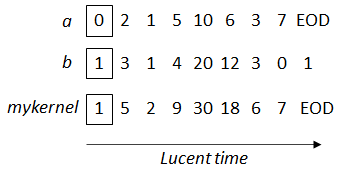}
\caption{Progression of stream values by time}
\label{fig:simple_addition}
\end{figure}

The progression of data streams \emph{a}, \emph{b}, and \emph{mykernel} are illustrated in Figure \ref{fig:simple_addition}, with the \emph{EOD} token used to signify End Of Data and terminate the filter. In this example \emph{mykernel} is a filter, with two inputs and an output, and is marked \emph{external} so it can be invoked directly from the host via the host-side Lucent API.

Listing \ref{lst:fby_conditionals} illustrates a slightly more complex example which contains conditionals and the followed by (\emph{fby} operator). We declare that the value of \emph{mykernel}, the filter's output stream, is 1 followed by 2 if the value of the \emph{a} stream at the specific time quanta is less than ten, otherwise the value streamed out is 0. The followed by (\emph{fby} operator) declares that a value is followed by another value next time quanta, for instance \emph{myseq:int = 0 fby 1 fby 2} defines the \emph{myseq} sequence (a sequence is internal to a filter) to be the values 0, 1, and 2 respectively as Lucent time progresses (and 2 infinitely after this point). It was Lucid that introduced the followed by operator, and this enables one to define iterative behaviour. At all points the programmer is explicitly considering the progression of time, the consumption of data and generation of results.


\begin{lstlisting}[frame=lines,caption={Code illustrating conditionals and followed by},label={lst:fby_conditionals}, numbers=left]
external mykernel:int(a:int) where:    
    mykernel=if (a < 10) then 1 fby 2 else 0 fi
\end{lstlisting}

Listing \ref{lst:1dstencil} illustrates a simple 1D stencil code performing one iteration of Laplace's equation for diffusion on a stream of (single precision) floating point numbers. There are three dataflow filters, \emph{mykernel} as the entry point, connecting the \emph{input} stream to the \emph{stencil} filter and this to the \emph{calc} filter which streams to the output of \emph{mykernel}. The \emph{stencil} filter is effectively a 1D shift buffer, with the programmer using a list to hold the three values of interest (the \emph{tl} operator takes the tail of the list, \emph{::} is list concatenation). The \emph{calc} filter defines an internal stream \emph{i} to be a counter (incremented each time quanta) and from the third time quanta onwards the expression at line 9 is evaluated and the result streamed out. 

\begin{lstlisting}[frame=lines,caption={1D stencil Lucent dataflow example},label={lst:1dstencil}, numbers=left]
filter stencil:list[float,3](in:float) where:
  stencil=if (in == EOD) then EOD else tl(stencil) :: [in] fi

filter calc:float(in:list[float,3]) where:
  i:int = 1 fby i+1
  calc=if (in == EOD) then:
          EOD
       else:
          if (i >= 2) then 0.5 * (at(in, 0) + at(in,2)) fi

external mykernel:float(input:float) where:
  mykernel=calc(stencil(input))
\end{lstlisting}

\section{Conclusions}
In this position paper we have described the benefits of embracing dataflow when writing high performance codes for reconfigurable architectures. By basing their abstract execution model on an application specific dataflow machine, then one can more readily write fast by construction codes for FPGAs. We have argued that modern takes on generations-old dataflow languages such as Lucid and Lustre are worth considering, and given a brief overview of our approach Lucent for HPC codes. Lucent itself contains much more functionality than we have had space to discuss, for instance including numerous types, selection of numeric precision and representation, multi-dimensional lists, the ability to direct which memory spaces data occupies (e.g. BRAM, URAM, DRAM, HBM), vectorised operations, user defined and generic types, and time dimensions. 



\bibliographystyle{abbrvnat}


\end{document}